\documentclass[11pt]{JHEP3}

\usepackage{epsfig}
\usepackage{graphicx}

\usepackage{amssymb}
\usepackage{citesort}
\usepackage{amsthm}

\usepackage{amsmath}
\usepackage{amsfonts}

\usepackage[active]{srcltx}

\newcommand{\To}{\longrightarrow}

\title{Classical conformal blocks
from TBA for the elliptic Calogero-Moser system}

\author{Marcin Pi\c{a}tek\\
Institute of Physics, University of Szczecin,
\\
ul. Wielkopolska 15, 70-451 Szczecin, Poland
\\ \\
Bogoliubov Laboratory of Theoretical Physics,
Joint Institute for Nuclear Research,
\\
141980 Dubna, Russia
\\ \\
E-mail: \email{piatek@fermi.fiz.univ.szczecin.pl}}

\abstract{The so-called Poghossian identities connecting the toric
and spherical blocks, the AGT relation on the torus and the
Nekrasov-Shatashvili formula for the elliptic Calogero-Moser Yang's
(eCMY) functional are used to derive certain expressions for the
classical 4-point block on the sphere. The main motivation for this
line of research is the longstanding open problem of uniformization
of the 4-punctured Riemann sphere, where the 4-point classical block
plays a crucial role. It is found that the obtained representation
for certain 4-point classical blocks implies the relation between
the accessory parameter of the Fuchsian uniformization of the
4-punctured sphere and the eCMY functional. Additionally, a relation
between the 4-point classical block and the $N_f=4$, ${\sf SU(2)}$
twisted superpotential is found and further used to re-derive the
instanton sector of the Seiberg-Witten prepotential of the $N_f=4$,
${\sf SU(2)}$ supersymmetric gauge theory from the classical block.
}

\begin{document}
\section{Introduction}
Let ${\cal C}_{g,n}$ be the Riemann surface with genus $g$ and
a set of points $z_1,\ldots,z_n$ removed from it. The basic objects of
any two-dimensional conformal field theory (2dCFT)
living on ${\cal C}_{g}$ \cite{Belavin:1984vu,EO}
are the $n$-point correlation functions of primary vertex operators
defined on ${\cal C}_{g,n}$.
Given a marking\footnote{A marking of the
Riemann surface ${\cal C}_{g,n}$ (for definition see \cite{JT0}) is a pants decomposition of
${\cal C}_{g,n}$ together with the corresponding trivalent graph.}
$\sigma$ of the Riemann surface ${\cal C}_{g,n}$
any correlation function can
be factorized according to the pattern given by a pants decomposition of
${\cal C}_{g,n}$ and written as a sum  (or an integral for theories with a continuous spectrum)
which includes the terms consisting of holomorphic and anti-holomorphic
conformal blocks times the 3-point functions of the model
for each pair of pants. The Virasoro conformal block
${\cal F}_{c,\alpha}^{(\sigma)}[\beta]({\sf z})$ on ${\cal C}_{g,n}$, where
$\alpha\equiv(\alpha_1, \ldots,\alpha_{3g-3+n})$,
$\beta\equiv(\beta_1,\ldots,\beta_n)$
depends on the cross ratios of the vertex operators locations denoted symbolically
by ${\sf z}$ and on the $3g-3+n$ intermediate conformal weights $\Delta_{\alpha_i}=
\alpha_i (Q -\alpha_i)$. Moreover, it depends on the $n$ external conformal weights
$\Delta_{\beta_a}=\beta_{a}(Q - \beta_a)$ and on the central charge $c$ which can be
parameterized as follows $c=1+6Q^2$ with
\begin{equation}\label{Q}
Q=b+b^{-1}.
\end{equation}

Conformal blocks are fully determined by the underlying conformal symmetry.
These functions possess an interesting, although not yet completely understood analytic
structure. In general, they can be expressed only as a formal power series and
no closed formula is known for its coefficients.
Among  the issues concerning
conformal blocks which are still not fully understood there is the problem
of their {\it semiclassical limit}.
This is the limit in which all parameters of the conformal blocks
tend to infinity in such a way that their ratios are fixed.
It is commonly believed that such limit exists and the conformal blocks behave in this limit
exponentially with respect to ${\sf z}$. This last has been however verified explicitly
only in the case of the conformal block on the 4-punctured sphere.
Indeed, the existence of the semiclassical limit of the 4-point Liouville correlation function
and the projection of this correlation function onto an appropriate intermediate conformal family
imply that the semiclassical limit of the 4-point conformal block \cite{Belavin:1984vu}
with ``heavy'' weights $\Delta=b^{-2}\delta$,
$\Delta_i=b^{-2}\delta_i,$ with $\delta,\delta_i = {\cal O}(1)$ has the following form:
\begin{equation}
\label{defccb}
{\cal F}_{\!c=1+6Q^2,\Delta}
\!\left[_{\Delta_{4}\;\Delta_{1}}^{\Delta_{3}\;\Delta_{2}}\right]
\!(x)
\; \stackrel{b\,\to\,0}{\,\To} \;
\exp \left\{
\frac{1}{b^2}\,{\sf f}_{\delta}
\!\left[_{\delta_{4}\;\delta_{1}}^{\delta_{3}\;\delta_{2}}\right]
\!(x)
\right\}.
\end{equation}
The function
${\sf f}_{\delta}\!\left[_{\delta_{4}\;\delta_{1}}^{\delta_{3}\;\delta_{2}}\right]\!(\,x)$
is  called the {\it classical conformal block} \cite{Zamolodchikov:1995aa} or
with some abuse of terms, the ``classical action'' \cite{Zam0,Zam}.
The existence of the semiclassical limit (\ref{defccb}) has been postulated first
in \cite{Zam0,Zam} where it has been pointed out that the
classical block is related to a certain monodromy problem
of a null vector decoupling equation in a similar
way in which the classical Liouville action is related to the
Fuchsian uniformization. This relation has been further used to
derive the $\Delta\to \infty$ limit of the 4-point conformal block
and its expansion in powers of the so-called elliptic variable.

Recently,  a considerable progress in the theory of conformal
blocks and their applications has been achieved.
This is mainly due to the discovery of the so-called AGT correspondence
by Alday, Gaiotto and Tachikawa in 2009 \cite{AGT}.
The AGT correspondence states that the Liouville field theory (LFT)
correlators on  ${\cal C}_{g,n}$
can be identified with the partition functions $Z_{{\cal T}_{g,n}}^{(\sigma)}$
of a class ${\cal T}_{g,n}$ (cf.\cite{G}) of four-dimensional
${\cal N}=2$ supersymmetric ${\sf SU}(2)$ quiver gauge theories\footnote{
Notice, that the Liouville theory correlators are modular \cite{HJSmodular} and crossing
symmetric \cite{JT1,JT2,PT1,PT2} and  do not depend on the choice of pants decomposition.
This strongly suggests the S-duality invariance of the gauge theory partition functions.}:
\begin{equation}\label{fullAGT}
\left\langle \prod\limits_{a=1}^{n}{\sf V}_{\beta_a}
\right\rangle_{{\cal C}_{g,n}}^{\textrm{LFT}} = Z_{{\cal T}_{g,n}}^{(\sigma)}.
\end{equation}

Let us describe the AGT proposal in more details. As well as the correlators of the Liouville theory,
also the partition functions $Z_{{\cal T}_{g,n}}^{(\sigma)}$ have an integral representation for
a fixed $\sigma$. Indeed, $Z_{{\cal T}_{g,n}}^{(\sigma)}$ can be written
as the integral over the holomorphic and anti-holomorphic Nekrasov
partition functions \cite{N,NekraOkun}:
\begin{equation}\label{IntegralRep}
Z_{{\cal T}_{g,n}}^{(\sigma)}=\int [da]\,
{\cal Z}_{{\sf Nekrasov}}^{(\sigma)}
\overline{{\cal Z}}_{{\sf Nekrasov}}^{(\sigma)},
\end{equation}
where $[da]$ is an appropriate measure. The Nekrasov partition function\footnote{
The dot in the first slot symbolizes
the dependence on the gauge
theory parameters, in general these are:
(i) the
$3g-3+n$ gluing parameters
${\sf q}=(q_i=\exp 2\pi \tau_i)_{i=1,\ldots,3g-3+n}$
associated with the pants decomposition, where $\tau_i =
\theta_{i}/2\pi + 4\pi i/g^{2}_{i}$
are complexified gauge couplings;
(ii) the $3g-3+n$ vacuum expectation values
(vev's) of the scalar fields in the vector multiplets
${\sf a}=(a_1, \ldots, a_{3g-3+n})$;
(iii) the $n$ mass parameters ${\sf m}=(m_1, \ldots,
m_n)$.
Finally, $\epsilon_1$, $\epsilon_2$
are the so-called
complex $\Omega$-background parameters
\cite{MNekraSha,ANekraSha,MNekraSha2}.}
$\mathcal{Z}_{{\sf Nekrasov}}(\,\cdot\,;\epsilon_1, \epsilon_2)$
appearing in (\ref{IntegralRep}) is a product of the three factors:
\begin{equation}
\label{FullNekrasov} {\cal Z}_{{\sf Nekrasov}}
= \mathcal{Z}_{{\sf class}}
\,\mathcal{Z}_{1-{\sf loop}}\, \mathcal{Z}_{{\sf inst}}.
\end{equation}
The first two factors $\mathcal{Z}_{{\sf class}}\,\mathcal{Z}_{1-{\sf loop}}
\equiv\mathcal{Z}_{{\sf pert}}$ describe the
contribution coming from perturbative calculations. Supersymmetry implies
that there are contributions to $\mathcal{Z}_{{\sf pert}}$ only at
tree- ($\mathcal{Z}_{{\sf class}}$) and 1-loop-level ($\mathcal{Z}_{1-{\sf loop}}$).
${\cal Z}_{{\sf inst}}$ is the instanton contribution.
A significant (Liouville independent, ``chiral'') part of the AGT conjecture is an exact correspondence
between the Virasoro conformal blocks ${\cal F}_{c,\alpha}^{(\sigma)}[\beta]({\sf z})$
on ${\cal C}_{g,n}$ and the instanton sectors
${\cal Z}_{{\sf inst}}$ of the Nekrasov partition functions of the gauge theories ${\cal T}_{g,n}$.

The AGT conjecture was actively studied starting
from the moment of its discovery. In particular,
various checks \cite{MMM,AlbaMoro} and proofs \cite{FL,MM2,HJS2} in special cases
have been investigated. Moreover, there have been attempts to explain
that conjecture from the points of view of
$M$-theory \cite{BT,ABT} and topological string theory
\cite{DV}. The direct consequences of AGT relations have been studied for instance in
\cite{Pogho,MM,MM3,MMM3,NX,T}.
Soon the AGT hypothesis has been extended to the $\sf
SU(N)$-gauge theories/$W_{\sf N}$ 2dCFT correspondence
\cite{Wyllard,MMU(3),MMMM}. Extensions to
asymptotically free theories \cite{G2,MMM2,AM}
and five-dimensional gauge theory
\cite{AY,AY2,Y2} have been analyzed too.
Additionally, this duality paved the way for new results in the theory of loop and
surface operators
\cite{AGGTV,DGOT,DGG,FP,KPW}
and has appeared in contexts related to matrix models
\cite{IO,MMM4,MSh,Sulkowski,TM,TM2}.
As previously mentioned, the studies of the AGT duality
have inspired several progresses inside pure 2dCFT, namely in
the theory of conformal blocks  \cite{Pogho,HJStorus,AlbaFatLitTar,BB}
as well as in Liouville theory \cite{HJSmodular}.

One of the immediate implications of the AGT proposal
is the passibility
of a derivation of the instanton contributions to the so-called Seiberg-Witten
{\it prepotentials} \cite{SW} from conformal blocks.
Let us recall that originally the Nekrasov partition functions
have been introduced to calculate the low energy effective gauge theories
prepotential $F\left(\,\cdot\,;\epsilon_1, \epsilon_2\right)$. For
$\epsilon_1, \epsilon_2\to 0$, the Nekrasov partition functions
behave as follows:
\begin{eqnarray}
\label{NekrasovAsym}
{\cal Z}_{{\sf Nekrasov}}(\,\cdot\,;\epsilon_1, \epsilon_2)&\To &
\exp\left\lbrace-\frac{1}{\epsilon_1 \epsilon_2}\;
F(\,\cdot\,;\epsilon_1, \epsilon_2)\right\rbrace,
\end{eqnarray}
where  $F\left(\,\cdot\,;\epsilon_1, \epsilon_2\right)$
becomes the Seiberg-Witten prepotential
$F\left(\,\cdot\,\right)$ in the limit $\epsilon_1,
\epsilon_2\to 0$.

The Nekrasov functions lead
also to an interesting application when one of  the
parameters $\epsilon_1, \epsilon_2$ is non-zero.
Recently, the limit $\epsilon_2\to 0$ (while $\epsilon_1$ is kept finite)
has been discussed in \cite{NekraSha}.
There has been observed that in this limit
the Nekrasov partition functions behave as
\begin{equation}\label{NSlimit}
{\cal Z}_{{\sf Nekrasov}}(\,\cdot\,;\epsilon_1, \epsilon_2)
\stackrel{\epsilon_2\to 0}{\To}
\exp\left\lbrace\frac{1}{\epsilon_2}\,
{\cal W}(\,\cdot\,;\epsilon_1)\right\rbrace,
\end{equation}
where ${\cal W}(\,\cdot\,;\epsilon_1)={\cal W}_{{\sf pert}}(\,\cdot\,;\epsilon_1)
+ {\cal W}_{{\sf inst}}(\,\cdot\,;\epsilon_1)$ is
the so-called effective {\it twisted superpotential}
of the corresponding two-dimensional gauge theories
restricted to the two-dimensional $\Omega$-background.
It is not difficult to realize, looking at particular examples
of AGT relations (see next Section),
that the limit $\epsilon_2 \to 0$ should correspond to the
classical limit $b\to 0$ of the conformal blocks.

On the other hand, twisted superpotentials play also a prominent role in another correspondence,
the so-called Bethe/gauge correspondence \cite{NekraSha,NekraSha2,NekraSha3}
which maps supersymmetric vacua of the ${\cal N}=2$ theories to Bethe states of quantum
integrable systems (QIS's). A result of that duality is that twisted superpotentials are identified with
Yang's functionals \cite{YY} which describe a spectrum of  QIS's. For instance,
in \cite{NekraSha} Nekrasov and Shatashvili have found that the twisted superpotential
$
{\cal W}^{{\cal N}=2^{\ast},{\sf U(N)}}(q,{\sf a},m;\epsilon_1)
$
serves as Yang's functional for the ${\sf N}$-particle elliptic Calogero-Moser (eCM)
system and for the periodic Toda chain (pToda).
Hance, by combining the AGT duality and the Bethe/gauge correspondence it is possible to link
classical blocks to Yang's functionals\footnote{Recently, such duality has been established in
\cite{Teschner} without employing the AGT conjecture.}.

The main aim of the present work is to exploit this link
to shed some light on the classical block on the sphere.
The motivation for this line of research is the longstanding
open problem of the uniformization of the 4-punctured sphere, where
the knowledge of the 4-point classical block plays a key role.

The structure of the paper is as follows.
In Section 2, quantum blocks on the torus and sphere and its so-called elliptic
representations are defined. Afterwards, the classical block on the sphere is introduced
and the relation between it and the instanton superpotential
of the ${\sf SU(2)}$ (or ${\sf U(2)}$), $N_f=4$ theory is written down.
This last relation is further used to re-derive in a consistent way the instanton contributions
to the Seiberg-Witten prepotential of the ${\sf SU(2)}$ (or ${\sf U(2)}$), $N_f=4$ theory
starting from the classical block.

In Section 3, using the so-called Poghossian identities, the AGT relation on the torus and
the Nekrasov-Shatashvili results from \cite{NekraSha}, the expressions for certain families of
classical blocks on the sphere are derived. It turns out that the most non-trivial parts
of the classical blocks are encoded in the closed formula of the
2-particle eCM Yang's functional \cite{NekraSha}.

In Section 4, the connection between classical Liouville
theory, semiclassical limit of quantum DOZZ Liouville theory and
the problem of uniformization of the 4-punctured Riemann sphere is recalled.
The role of the classical block on the sphere in that context is clarified.
Then, the expressions of certain classical blocks derived in Section 3 are
applied to calculate the accessory parameter of the Fuchsian uniformization of the
4-punctured sphere. The relation between the accessory parameter
and the 2-particle eCM Yang's functional is found.

Finally, Section 5 contains conclusions and a discussion of the open
problems and possible extensions of the present work.

\section{Classical AGT relation on a sphere}
\subsection{Quantum blocks on torus and sphere}
Let $\hat q=\textrm{e}^{2\pi i \tau}$  be the elliptic variable on the torus with
modular parameter $\tau$ and $x$ be the moduli of the 4-punctured sphere.
The quantum conformal blocks on the 1-punctured torus and on the 4-punctured sphere
are defined as formal power series:
\begin{eqnarray}
\label{torusblock} {\cal
F}_{c,\Delta}^{\tilde\Delta}(q)&=&q^{\Delta-\frac{c}{24}}
\left(1+\sum\limits_{n=1}^{\infty}{\cal
F}_{c,\Delta}^{\tilde\Delta,n}q^n \right),
\\
\label{torusBlockCoeff} \mathcal{F}^{\tilde\Delta,
n}_{c,\Delta}&=&\sum\limits_{n=|I|=|J|}
\left\langle\nu_{\Delta,I},V_{\tilde\Delta}(1)\nu_{\Delta,J}\right\rangle
\;\Big[ G_{c,\Delta}\Big]^{IJ},
\end{eqnarray}
and
\begin{eqnarray}
\label{block} {\cal
F}_{c,\Delta}\!\left[_{\Delta_{4}\;\Delta_{1}}^{\Delta_{3}\;\Delta_{2}}\right]\!(\,x)
&=& x^{\Delta-\Delta_{2}-\Delta_{1}}\left( 1 + \sum_{n=1}^\infty
{\cal
F}^{\,n}_{c,\Delta}\!\left[_{\Delta_{4}\;\Delta_{1}}^{\Delta_{3}\;\Delta_{2}}\right]
x^{\,n} \right),
\\
{\cal
F}^{\,n}_{c,\Delta}\!\left[_{\Delta_{4}\;\Delta_{1}}^{\Delta_{3}\;\Delta_{2}}\right]
&=& \label{blockCeef} \sum\limits_{n=|I|=|J|}
\left\langle\nu_{\Delta_4},V_{\Delta_3}(1)\nu_{\Delta,I}\right\rangle
\;
\Big[G_{c,\Delta}\Big]^{IJ}
\;
\left\langle\nu_{\Delta,J},V_{\Delta_2}(1)\nu_{\Delta_1}\right\rangle
\end{eqnarray}
respectively. In the above equations
$\Big[ G_{c,\Delta}\Big]^{IJ}$ is the
inverse of the Gram matrix $ \Big[
G_{c,\Delta}\Big]_{IJ}=\langle \nu_{\Delta,I},\nu_{\Delta,J} \rangle
$ of the standard symmetric bilinear form in the Verma module ${\cal
V}_{\Delta}=\bigoplus_{n=0}^{\infty}{\cal V}_{\Delta}^{n}$,
\begin{eqnarray*}
{\cal V}_{\Delta}^{n}=\textrm{Span}\Big\lbrace
\nu^n_{\Delta,I}=L_{-I}\nu_{\Delta} = L_{-i_{k}}\ldots
L_{-i_{2}}L_{-i_{1}} \nu_\Delta
&:&\\
&&\hspace{-200pt} I=( i_{k}\geq \ldots\geq i_{1}\geq 1)
\;\textrm{an}\; \textrm{ordered} \;
\textrm{set}\; \textrm{of }\;\textrm{positive} \;\textrm{integers} \;\\
&&\hspace{-100pt} \textrm{of}\;
\textrm{the}\;\textrm{length}\;|I|\equiv i_{1}+\ldots+i_{k}=n
\Big\rbrace.
\end{eqnarray*}
The operator $V_{\Delta}$ in the matrix elements is the normalized
primary chiral vertex operator acting between the Verma modules
$$
\left\langle\nu_{\Delta_i},V_{\Delta_{j}}(z)\nu_{\Delta_k}\right\rangle
=z^{\Delta_i -\Delta_{j}-\Delta_k}.
$$
In order to calculate the matrix elements in (\ref{torusBlockCoeff}), (\ref{blockCeef})
it is enough to know the covariance properties of the primary chiral
vertex operator with respect to the Virasoro algebra:
$$
\left[L_n , V_{\Delta}(z)\right] = z^{n}\left(z
\frac{d}{dz} + (n+1)\Delta
\right)V_{\Delta}(z)\,,\;\;\;\;\;\;\;\;\;\;n\in\mathbb{Z}.
$$

As the dimension of ${\cal V}^n_\Delta$ grows rapidly with $n$,
the calculations of conformal block coefficients by inverting
the Gram matrices become very laborious for higher orders.
A more efficient method based on recurrence relations for the coefficients
can be used \cite{Zam,FL,Pogho,HJStorus,Zamolodchikov:ie}.

It is convenient to introduce the so-called elliptic conformal blocks.
The 1-point elliptic block ${\cal H}_{c,\Delta}^{\tilde\Delta}(\hat q)$ on the torus with
modular parameter $\tau$ and elliptic variable $\hat q=\textrm{e}^{2\pi i \tau}$
is defined by \cite{FL,HJStorus}:
\begin{equation}\label{ellipticTorus}
{\cal H}_{c,\Delta}^{\tilde\Delta}(\hat q)=
{\hat q\,}^{\frac{c-1}{24}-\Delta}\,
\eta(\hat q)\,{\cal F}_{c,\Delta}^{\tilde\Delta}(\hat q)=
\sum\limits_{n=0}^{\infty}\hat q^{\,n}{\cal H}_{c,\Delta}^{\tilde\Delta,n},
\end{equation}
where $\eta(\hat q)={\hat q}^{\frac{1}{24}}\prod_{n=1}^{\infty}(1-{\hat q}^{\,n})$ is the Dedekind eta function.

The 4-point elliptic block on the sphere ${\cal H}_{\!c,\Delta}
\!\left[_{\Delta_{4}\;\Delta_{1}}^{\Delta_{3}\;\Delta_{2}}\right]\!(q)$,
where
\begin{equation}\label{nome}
q\equiv q(x)={\rm e}^{-\pi {K(1-x)\over K(x)}},
\;\;\;\;\;\;\;\;\;\;\;\;\;\;
K(x)=\int\limits_{0}^1{dt\over \sqrt{(1-t^2)(1-xt^2)}}
\end{equation}
is defined by \cite{Zam}
\begin{eqnarray}
\label{Zamolo}
{\cal F}_{\!c,\Delta}\!\left[_{\Delta_{4}\;\Delta_{1}}^{\Delta_{3}\;\Delta_{2}}\right]\!(\,x)
&=&
x ^{{c-1\over 24}-\Delta_1-\Delta_2}
(1-x)^{{c-1\over 24}-\Delta_2-\Delta_3}\nonumber
\\
&\times &
\theta_3(q)^{{c-1\over 2}-4(\Delta_1+\Delta_2+\Delta_3+\Delta_4)}
\;
(16 q)^{\Delta - {c-1\over 24}}\,
{\cal H}_{\!c,\Delta}\!\left[_{\Delta_{4}\;\Delta_{1}}^{\Delta_{3}\;\Delta_{2}}\right]\!(\,q),
\end{eqnarray}
where
\begin{equation}
\label{HH}
{\cal H}_{\!c,\Delta}\!\left[_{\Delta_{4}\;\Delta_{1}}^{\Delta_{3}\;\Delta_{2}}\right]\!(\,q)
=  1 + \sum_{n=1}^\infty (16q)^{\;n}
{\cal H}^{\,n}_{\!c,\Delta}\!\left[_{\Delta_{4}\;\Delta_{1}}^{\Delta_{3}\;\Delta_{2}}\right].
\end{equation}
The coefficients in (\ref{HH}) are uniquely determined by the
recursion relation\footnote{
Here, for the central charge parameterized as
$c=1+6(b+{\textstyle\frac{1}{b}})^2\equiv1+6Q^2$ the degenerate
conformal weights are of the form
$
\Delta_{rs}(c )={1-r^2\over 4}\,b^2 +{1-rs\over 2} +{1-s^2\over
4} {1\over b^2}\ .
$
The explicit form of the coefficients
$R^{\,rs}_c\!\left[_{\Delta_{4}\;\Delta_{1}}^{\Delta_{3}\;\Delta_{2}}\right]$
is known \cite{ZZ}.
}:
\begin{eqnarray}
\label{recrelHH}
{\cal H}^{\,n}_{\!c,\,\Delta}\!\left[_{\Delta_{4}\;\Delta_{1}}^{\Delta_{3}\;\Delta_{2}}\right]
&=&
\sum_{\begin{array}{c}\scriptstyle r\geq 1\; s\geq 1\\[-3pt]\scriptstyle n\,\geq \,rs \,\geq 1 \end{array}}
\!\!\!\!\! \frac{
R^{\,rs}_c\!\left[_{\Delta_{4}\;\Delta_{1}}^{\Delta_{3}\;\Delta_{2}}\right]}
{\Delta-\Delta_{rs}(c)} \;{\cal H}^{\,n-rs}_{\!c,\,\Delta_{rs}(c) +rs}\!
\left[_{\Delta_{4}\;\Delta_{1}}^{\Delta_{3}\;\Delta_{2}}\right],
\hskip 1cm n>0.
\end{eqnarray}

The toric and spherical blocks appear in the best known examples of the ``chiral''
AGT relations.
\begin{enumerate}
\item
${\cal F}_{c,\Delta}^{\tilde\Delta}(\hat q)$ is identified with the Nekrasov instanton
partition function of the ${\cal N}=2^{\ast}$, ${\sf SU(2)}$ gauge
theory (which equals to
$
[{\cal Z}_{\sf inst}^{{\sf U(1)}}]^{-1}
\times {\cal Z}_{\sf inst}^{{\sf U(2)}}
$
as it is written in the second line of the equation below):
\begin{eqnarray}
\label{AGTtorus}
{\hat q\,}^{\frac{c}{24}-\Delta}
{\cal F}_{c,\Delta}^{\tilde\Delta}(\hat q)&=&
\,
\mathcal{Z}^{{\cal N}=2^{\ast},{\sf SU(2)}}_{{\sf
inst}}(\hat q,a,m;\epsilon_1, \epsilon_2)\nonumber
\\
&=&\hat\eta(\hat q)^{1-2\tilde\Delta}
\,\mathcal{Z}^{{\cal N}=2^{\ast},{\sf U(2)}}_{{\sf
inst}}(\hat q,a,m;\epsilon_1, \epsilon_2).
\end{eqnarray}
In eq. (\ref{AGTtorus})
$\hat\eta(\hat q)\equiv\prod_{n=1}^{\infty}(1-\hat q^{\,n})$ and
the torus block parameters, namely the external conformal weight
$\tilde\Delta$, the intermediate weight $\Delta$ and the Virasoro
central charge $c$ can be expressed in terms of the ${\cal N}=2^{\ast}$, ${\sf SU(2)}$
super-Yang-Mills theory parameters as follows
\begin{equation}\label{paraTorus}
\tilde\Delta=-\frac{m(m+\epsilon_1+\epsilon_2)}{\epsilon_1
\epsilon_2}, \;\;\;\;\;
\Delta=\frac{(\epsilon_1+\epsilon_2)^2-4a^2}{4\epsilon_1\epsilon_2},
\;\;\;\;\;
c=1+6\frac{(\epsilon_1+\epsilon_2)^2}{\epsilon_1\epsilon_2} \equiv
1+6Q^2.
\end{equation}
In eq. (\ref{paraTorus}) $m$ is the mass of the adjoint
hypermultiplet, $a$ is the vacuum expectation value (vev) of the complex scalar of the gauge
multiplet and $\epsilon_1, \epsilon_2$ are $\Omega-$background
parameters.
\item
${\cal F}_{c,\Delta}\!\left[_{\Delta_{4}\;\Delta_{1}}^{\Delta_{3}\;\Delta_{2}}\right]\!(\,x)$
is related to the Nekrasov instanton partition function of the ${\cal N}=2$, ${\sf SU(2)}$ (or ${\sf U(2)}$)
super-Yang-Mills theory with four flavors ($N_f = 4$):
\begin{eqnarray}\label{AGTsphere}
x^{\Delta_1 + \Delta_2 -\Delta}\,
{\cal F}_{c,\Delta}\!\left[_{\Delta_{4}\;\Delta_{1}}^{\Delta_{3}\;\Delta_{2}}\right]\!(\,x)
&=&
\mathcal{Z}^{N_f = 4,{\sf SU(2)}}_{{\sf inst}}(x,a,\mu_i;\epsilon_1, \epsilon_2)
\\
&=& (1-x)^{-\frac{(\mu_1+\mu_2)(\mu_3+\mu_4)}{2\epsilon_1\epsilon_2}}\,
\mathcal{Z}^{N_f = 4,{\sf U(2)}}_{{\sf inst}}(x,a,\mu_i;\epsilon_1, \epsilon_2).\nonumber
\end{eqnarray}
In eq. (\ref{AGTsphere}) the central charge $c$, the intermediate weight $\Delta$
and the four external weights $\Delta_i$ are related to
the vev $a$, the masses $\mu_1,\ldots,\mu_4$ of  the four
hypermultiplets and  the  $\Omega$-background parameters
$\epsilon_1, \epsilon_2$ respectively by the following relations:
\begin{eqnarray}
\label{external}
c=1+6\frac{(\epsilon_1+\epsilon_2)^2}{\epsilon_1\epsilon_2}
\equiv 1+6Q^2,&\;\;\;\;\;&
\Delta=\frac{(\epsilon_1+\epsilon_2)^2-4a^2}{4\epsilon_1\epsilon_2},\\
\Delta_1=
\frac{\frac{1}{4}(\epsilon_1\!+\!\epsilon_2)^2-\frac{1}{4}(\mu_1\! - \!\mu_2)^2}{\epsilon_1\epsilon_2},
&\;\;\;\;\;&
\Delta_2=
\frac{\frac{1}{2}(\mu_1\!
+\! \mu_2)(\epsilon_1\!+\!\epsilon_2
-\frac{1}{2}(\mu_1\! +\! \mu_2))}{\epsilon_1\epsilon_2},\nonumber
\\
\Delta_3=
\frac{\frac{1}{2}(\mu_3\! +\! \mu_4)
(\epsilon_1\!+\!\epsilon_2\!-\!\frac{1}{2}(\mu_3 \!+\! \mu_4))}{\epsilon_1\epsilon_2}
,\nonumber
&\;\;\;\;\;&
\Delta_4=
\frac{\frac{1}{4}(\epsilon_1\!+\!\epsilon_2)^2\!-\!\frac{1}{4}(\mu_3\! -\! \mu_4)^2}{\epsilon_1\epsilon_2}.
\nonumber
\end{eqnarray}
\end{enumerate}
The identities (\ref{AGTtorus}) and (\ref{AGTsphere})
are understood as equalities between the coefficients of the expansions
of both sides in powers of $\hat q$ and $x$ respectively\footnote{
For definition of the expansion coefficients of
$\mathcal{Z}^{{\cal N}=2^{\ast},{\sf U(2)}}_{{\sf
inst}}$ and $\mathcal{Z}^{N_f = 4,{\sf U(2)}}_{{\sf inst}}$
see \cite{NekraSha,N}.}.
In both the above examples the background charge is given by
$$
Q={\sqrt{\frac{\epsilon_2}{\epsilon_1}}}
+\sqrt{\frac{\epsilon_1}{\epsilon_2}}.
$$
In order to be consistent with the standard parametrization
(\ref{Q}) we shall set:
\begin{equation}\label{relation}
{\sqrt{\frac{\epsilon_2}{\epsilon_1}}}=b.
\end{equation}
\subsection{Seiberg-Witten prepotential from the classical 4-point block}
The asymptotic behavior (\ref{defccb}) implies the following expansion of the 4-point classical block:
\begin{eqnarray}
\label{classblock}
{\sf f}_{\delta}\!\left[_{\delta_{4}\;\delta_{1}}^{\delta_{3}\;\delta_{2}}\right]\!(\,x)
&=& (\delta-\delta_1-\delta_2) \log x +  \sum_{n=1}^\infty
x^{n}\, {\sf f}^{\,n}_{\delta}\!\left[_{\delta_{4}\;\delta_{1}}^{\delta_{3}\;\delta_{2}}\right]
\nonumber\\
&=& (\delta-\delta_1-\delta_2) \log x + \frac{(\delta + \delta_3 -
\delta_4)(\delta + \delta_2 - \delta_1)}{2\delta}\;x +\ldots\;.
\end{eqnarray}
The coefficients
${\sf f}^{\,n}_{\delta}\!\left[_{\delta_{4}\;\delta_{1}}^{\delta_{3}\;\delta_{2}}\right]$
in (\ref{classblock})
are calculated directly from the limit (\ref{defccb}) and the power expansion
of the quantum block:
\begin{eqnarray*}
\label{limit} \sum_{n=1}^\infty x^{\;n}
{\sf f}^{\,n}_{\delta}\!\left[_{\delta_{4}\;\delta_{1}}^{\delta_{3}\;\delta_{2}}\right]
&=& \lim\limits_{b \to 0} {b^2} \log\left(1 +
\sum_{n=1}^\infty {\cal
F}^{\,n}_{\!c,\Delta}\!\left[_{\Delta_{4}\;\Delta_{1}}^{\Delta_{3}\;\Delta_{2}}\right]
x^{\;n} \right)
\\
&=&
\lim\limits_{b \to 0} {b^2} \log\left(1 +
\frac{(\Delta+\Delta_3-\Delta_4)(\Delta+\Delta_2-\Delta_1)}{2\Delta}\; x^{\;n}
+\ldots\right),
\end{eqnarray*}
where on the r.h.s. one first expands the logarithm into
power series and then the limit is taken of each term separately.

On the other hand the representation (\ref{Zamolo}) and the asymptotic behavior (\ref{defccb})
imply another representation for the classical 4-point block on the sphere:
\begin{eqnarray}
\label{Hclassblock}
{\sf f}_{\delta}\!\left[_{\delta_{4}\;\delta_{1}}^{\delta_{3}\;\delta_{2}}\right]\!(\,x)
&=&
 ({\textstyle {1\over 4}}-\delta_1-\delta_2) \log x
+
 ({\textstyle {1\over 4}}-\delta_2-\delta_3) \log (1-x)\\
 \nonumber
&+&
 (3-4(\delta_1+\delta_2+\delta_3+\delta_4)) \log
 \left(\theta_3(q(x))\right)\\
 \nonumber
 &+& (\delta -{\textstyle {1\over 4}}) \log (16q(x))
+
{\sf h}_{\delta}\!\left[_{\delta_{4}\;\delta_{1}}^{\delta_{3}\;\delta_{2}}\right](q(x)),
\end{eqnarray}
where the elliptic classical block
${\sf h}_{\delta}\!\left[_{\delta_{4}\;\delta_{1}}^{\delta_{3}\;\delta_{2}}\right](q)$
is defined by the expansion
\begin{equation}
\label{hexp}
{\sf h}_{\delta}\!\left[_{\delta_{4}\;\delta_{1}}^{\delta_{3}\;\delta_{2}}\right](q)
 =\sum_{n=1}^\infty
(16q)^{\;n} {\sf h}^{\,n}_{\delta}\!\left[_{\delta_{4}\;\delta_{1}}^{\delta_{3}\;\delta_{2}}\right].
\end{equation}
The coefficients ${\sf h}^{\,n}_{\delta}\!\left[_{\delta_{4}\;\delta_{1}}^{\delta_{3}\;\delta_{2}}\right]$
are obtained step by step from the limit (\ref{defccb}):
\begin{equation}
\label{hlimit}
\sum_{n=1}^\infty
(16q)^{\;n} {\sf h}^{\,n}_{\delta}\!\left[_{\delta_{4}\;\delta_{1}}^{\delta_{3}\;\delta_{2}}\right] =
\lim\limits_{b \to 0} {b^2} \log\left(1 + \sum_{n=1}^\infty
(16q)^{\;n} {\cal H}^{\,n}_{\!c,\Delta}\!\left[_{\Delta_{4}\;\Delta_{1}}^{\Delta_{3}\;\Delta_{2}}\right]
\right).
\end{equation}

As mentioned in the Introduction classical blocks correspond to twisted superpotentials.
In the case under consideration, joining together (\ref{defccb}), (\ref{NSlimit}), (\ref{AGTsphere}) and
(\ref{relation}),  one gets
\begin{eqnarray}
\label{ClassAGTsphere}
\left(\delta_{1}+\delta_{2}-\delta\right)\log x + {\sf f}_{\delta}
\!\left[_{\delta_{4}\;\delta_{1}}^{\delta_{3}\;\delta_{2}}\right](x)
&=&
\frac{1}{\epsilon_1}\;
{\cal W}_{{\sf inst}}^{N_f=4,{\sf SU(2)}}(x,a,\mu_i;\epsilon_1)
\\
&&\hspace{-130pt}=\;\frac{1}{\epsilon_1}\;
{\cal W}_{{\sf inst}}^{N_f=4,{\sf U(2)}}(x,a,\mu_i;\epsilon_1)
-\frac{(\mu_1+\mu_2)(\mu_3+\mu_4)}{2\epsilon_{1}^{2}}
\;\log(1-x),\nonumber
\end{eqnarray}
where the classical weights are defined by
$\delta\equiv\lim_{b\to 0}b^2 \Delta
=\lim_{\epsilon_2\to 0}\frac{\epsilon_2}{\epsilon_1}\,\Delta$
and
$\delta_i\equiv\lim_{b\to 0}b^2 \Delta_i
=\lim_{\epsilon_2\to 0}\frac{\epsilon_2}{\epsilon_1}\,\Delta_i$.
Explicitly, they read as follows
\begin{eqnarray}
\label{classW}
\delta&=&\frac{1}{4}-\frac{a^2}{\epsilon^{2}_{1}},\;\;\;\;\;\;\;\;\;
\delta_{1}\;=\;\frac{1}{4}-\frac{(\mu_1 - \mu_2)^2}{4\epsilon^{2}_{1}}
,\;\;\;\;\;\;\;\;\;
\delta_{2}\;=\;\frac{\mu_1 + \mu_2}{2\epsilon_1}-\frac{(\mu_1 +
\mu_2)^2}{4\epsilon^{2}_{1}},
\nonumber\\[3pt]
\delta_{3}&=&\frac{\mu_3 + \mu_4}{2\epsilon_1}-\frac{(\mu_3 +
\mu_4)^2}{4\epsilon^{2}_{1}},\;\;\;\;\;\;\;\;\;
\delta_{4}\;=\;\frac{1}{4}-\frac{(\mu_3 - \mu_4)^2}{4\epsilon^{2}_{1}}.
\end{eqnarray}

The relation (\ref{ClassAGTsphere}) gives rise to a self-consistent method to calculate
from the classical block the instanton prepotential
for the ${\sf SU(2)}$ (or ${\sf U(2)}$), $N_f = 4$ theory
(cf.\cite{Tai}). First, notice that from (\ref{NekrasovAsym})
and (\ref{NSlimit}) one gets
\begin{equation}
\label{SW} {\cal W}(\,\cdot\,;\epsilon_1)\; \stackrel{\epsilon_1\,\to\,0}{\,\To} \; -
\frac{F\left(\,\cdot\,\right)}{\epsilon_1}\,.
\end{equation}
Then, using the above formula and eq.
(\ref{ClassAGTsphere}) with the values of the classical weights expressed in
(\ref{classW}) one can write the following relation in the case of the ${\sf SU(2)}$ theory:
\begin{eqnarray}\label{SWSu2}
F_{{\sf inst}}^{N_f=4,{\sf SU(2)}}(x,a,\mu_i)&=& - \lim\limits_{\epsilon_1\to 0}\epsilon_1
\,{\cal W}_{{\sf inst}}^{N_f=4,{\sf SU(2)}}(x,a,\mu_i;\epsilon_1)\nonumber\\
&=& -\lim\limits_{\epsilon_1\to 0}\left(
\epsilon_{1}^{2}
\,{\sf f}_{\delta}
\!\left[_{\delta_{4}\;\delta_{1}}^{\delta_{3}\;\delta_{2}}\right](x)
+\epsilon_{1}^{2}\,
(\delta_1 +\delta_2 -\delta)\log x\right).
\end{eqnarray}
Analogously, for the ${\sf U(2)}$ theory one gets:
\begin{eqnarray}\label{SWp}
F_{{\sf inst}}^{N_f=4,{\sf U(2)}}(x,a,\mu_i)&=& -\lim\limits_{\epsilon_1\to 0}\epsilon_1
\,{\cal W}_{{\sf inst}}^{N_f=4,{\sf U(2)}}(x,a,\mu_i;\epsilon_1)\nonumber
\\
&=&
-\lim\limits_{\epsilon_1\to 0}\left(
\epsilon_{1}^{2}
\,{\sf f}_{\delta}
\!\left[_{\delta_{4}\;\delta_{1}}^{\delta_{3}\;\delta_{2}}\right](x)
+\epsilon_{1}^{2}\,
(\delta_1 +\delta_2 -\delta)\log x\right)\nonumber
\\
&-&\frac{1}{2}(\mu_1+\mu_2)(\mu_3+\mu_4)\log(1-x).
\end{eqnarray}
Eq. (\ref{SWSu2}) and (\ref{SWp}) imply that:
\begin{equation}\label{ab}
F_{{\sf inst}}^{N_f=4,{\sf U(2)}}(x,a,\mu_i)=F_{{\sf inst}}^{N_f=4,{\sf SU(2)}}(x,a,\mu_i)
-\frac{1}{2}(\mu_1+\mu_2)(\mu_3+\mu_4)\log(1-x).
\end{equation}
The expansion of the instanton ${\sf SU(2)}$ prepotential can be obtained from eq. (\ref{SWSu2})
after using the representation (\ref{classblock}) of the classical block. The result is the expansion
in the bare coupling $x$:
\begin{eqnarray}\label{prepox}
F_{{\sf inst}}^{N_f=4,{\sf SU(2)}}(x,a,\mu_i)&=&
\frac{a^4+a^2(\mu_1\mu_2 + \mu_3\mu_4)+\mu_1\mu_2\mu_3\mu_4}{2 a^2}\,x\\
&&\hspace{-40pt}+\;
\Big[\frac{1}{64 a^6}\Big( 13a^8+5(\mu_1 \mu_2 \mu_3 \mu_4)^2\nonumber
\\
&&\hspace{-40pt}+\;
a^6(\mu_{1}^{2}+\mu_{2}^{2}+\mu_{3}^{2}+\mu_{4}^{2}
+ 16\mu_1\mu_2+16\mu_3\mu_4)\nonumber
\\
&&\hspace{-40pt}+\;
a^4 (16\mu_1 \mu_2 \mu_3 \mu_4 +(\mu_3 \mu_4)^2 + (\mu_2 \mu_3)^2 +(\mu_2 \mu_4)^2 +\mu_{1}^{2}
(\mu_{2}^{2}+\mu_{3}^{2}+\mu_{4}^{2}))\nonumber
\\
&&\hspace{-40pt}-\;
3a^2 ((\mu_2 \mu_3 \mu_4)^2 + (\mu_1 \mu_3 \mu_4)^2 + (\mu_1 \mu_2 \mu_3)^2+(\mu_1 \mu_2 \mu_4)^2)
\Big)\Big]\;x^2+\ldots\,.\nonumber
\end{eqnarray}
Analogously, an expansion of the instanton ${\sf SU(2)}$ prepotential with respect to the ``renormalized''
coupling $q(x)$ may be derived from eq. (\ref{SWSu2}) after applying the representation (\ref{Hclassblock})
of the conformal block:
\begin{eqnarray}\label{prepoq}
F_{{\sf inst}}^{N_f=4,{\sf SU(2)}}(x,a,\mu_i)&=&
a^2\log\left(\frac{16q(x)}{x}\right)
-\frac{1}{4}\,\left((\mu_1+\mu_2)^2 + (\mu_3+\mu_4)^2\right)\log(1-x)\nonumber
\\
&-& \sum_{i=1}^{4}\mu_{i}^{2}\log\left(
{\textstyle{2\over\pi}}K(x)\right) - \lim\limits_{\epsilon_1 \to
0}\epsilon_{1}^{2}\;
{\sf h}_{\delta}\!\left[_{\delta_{4}\;\delta_{1}}^{\delta_{3}\;\delta_{2}}\right]\!(q(x)),
\end{eqnarray}
where for example up to $q^2$:
\begin{eqnarray}\label{limith}
- \lim\limits_{\epsilon_1 \to
0}\epsilon_{1}^{2}\;
{\sf h}_{\delta}\!\left[_{\delta_{4}\;\delta_{1}}^{\delta_{3}\;\delta_{2}}\right]\!(q)
&=&
8a^{-2}\mu_1\mu_2\mu_3\mu_4\,q\nonumber\\
&+&\Big(
4a^{-2}\Big[(\mu_3\mu_4)^2+(\mu_2\mu_3)^2+(\mu_2\mu_4)^2+\mu_{1}^{2}(\mu_{2}^{2}+\mu_{3}^{2}+\mu_{4}^{2})\Big]\nonumber
\\
&-&12a^{-4}\Big[(\mu_2\mu_3\mu_4)^2+\mu_{1}^{2}((\mu_3\mu_4)^2+\mu_{2}^{2}(\mu_{3}^{2}+\mu_{4}^{2}))\Big]\nonumber
\\
&+&20a^{-6}(\mu_1\mu_2\mu_3\mu_4)^2
\Big)\,q^2+\ldots\;.
\end{eqnarray}

Expansion (\ref{prepox}) agrees with the formula obtained from the instanton
partition function.
To show the consistency of our approach, we plug in the expansion (\ref{prepox}) in eq. (\ref{ab}). The result reproduces
in the case of equal masses the expansion of the instanton ${\sf U(2)}$ prepotential of ref. \cite{TM2}:
\begin{eqnarray*}
F_{{\sf inst}}^{N_f=4,{\sf U(2)}}(x,a,\mu_i=\mu)&=& \frac{a^4+6a^2 \mu^2 +\mu^4}{2a^2}\,x
\\
&+&
\frac{13a^8+100 a^6 \mu^2 + 22 a^4 \mu^4 - 12 a^2 \mu^6 + 5\mu^8}{64 a^6}\,x^2 + \ldots\,.
\end{eqnarray*}

Let us close this section by remarking that in the intermediate steps of the calculation
of eq. (\ref{prepoq}) it
appears also the following important formula for the $N_f=4$, ${\sf SU(2)}$ twisted superpotential:
\begin{eqnarray}
\label{Well} {\cal W}_{{\sf inst}}^{N_f=4,{\sf SU(2)}}(x,a,\mu_i;\epsilon_1)
&=&
-\frac{a^2}{\epsilon_1}\log\left(\frac{16q(x)}{x}\right)\\
&+&
\left[\frac{\epsilon_1}{4}+\frac{1}{4\epsilon_1}\,\left((\mu_1+\mu_2)^2 + (\mu_3+\mu_4)^2\right)\right]\log(1-x)\nonumber
\\
&+& \left[\frac{\epsilon_1}{2} - \sum_{i=1}^{4}\mu_i
+\frac{1}{\epsilon_1}\sum_{i=1}^{4}\mu_{i}^{2}\right] \log\left(
{\textstyle{2\over\pi}}K(q)\right) + \epsilon_1\,
{\sf h}_{\delta}\!\left[_{\delta_{4}\;\delta_{1}}^{\delta_{3}\;\delta_{2}}\right]\!(q(x)).
\nonumber
\end{eqnarray}
In \cite{MMM3} the r.h.s. of eq. (\ref{prepoq}) and more precisely its limit
for large values of $a$ has been interpreted as the instanton prepotential of the four-dimensional
$N_f=4$, ${\sf SU(2)}$ theory written in the renormalized coupling $q(x)$.
Probably the same interpretation is valid in the case of the
two-dimensional theory determined by ${\cal W}_{{\sf inst}}^{N_f=4,{\sf SU(2)}}$.

\section{Four-point classical blocks from TBA for the eCM system}
Let us come back to quantum blocks for a while.
There exist amazing relationships between conformal blocks on the torus and
sphere known as Poghossian identities \cite{Pogho}.
In order to spell out these relations quickly and clearly
let us modify a little bit the notation and denote by
${\cal H}_{c,\Delta_{\lambda}}^{\tilde\lambda}(q)$
the torus elliptic block and by
$
{\cal H}_{\!c,\Delta_{\lambda}}\!\left[_{\lambda_{4}\;\lambda_{1}}
^{\lambda_{3}\;\lambda_{2}}\right]\!(\,q)
$
the elliptic block on the sphere, where the parameters $\lambda$, $\tilde\lambda$, $\lambda_i$
are related to the conformal weights in the following way:
\begin{equation}
\label{wag}
\Delta\equiv\Delta_{\lambda}={\textstyle\frac{1}{4}}\left(Q^2-\lambda^2\right),
\;\;\;\;\;
\tilde\Delta\equiv\Delta_{\tilde\lambda}={\textstyle\frac{1}{4}}\left(Q^2-\tilde\lambda^2\right),
\;\;\;\;\;
\Delta_i\equiv\Delta_{\lambda_i}={\textstyle\frac{1}{4}}\left(Q^2-\lambda^{2}_{i}\right).
\end{equation}
Additionally, the following parametrization for the central charge is assumed:
$c=1+6(b+{\textstyle\frac{1}{b}})^2\equiv1+6Q^2$.
In \cite{HJStorus} it has been proved that
\begin{eqnarray}\label{Pogho1}
{\cal H}_{c,\Delta_{\lambda}}^{\tilde\lambda}(q) &=&
{\cal H}_{\!c,\Delta_{\lambda}}\!\left[_{\;\frac{b}{2}-\frac{1}{2b}\;
\;\;\;\frac{Q}{2}\;}
^{\;\frac{\tilde\lambda}{2}\;\;\;\;\;\;\;\;
\;\;\;\frac{\tilde\lambda}{2}}\right]\!(\,q)
\nonumber\\[3pt]
&=&
{\cal H}_{\!c,\Delta_{\lambda}}\!\left[_{\;\frac{1}{2b}\;
\;\;\;\;\;\;\;\;\;\;\;\frac{1}{2b}\;}
^{\;\frac{\tilde\lambda}{2}-\frac{b}{2}\;\;\;\;
\frac{\tilde\lambda}{2}+\frac{b}{2}\;}\right]\!(\,q)
\nonumber\\[3pt]
&=&
{\cal H}_{\!c,\Delta_{\lambda}}\!\left[_{\;\frac{b}{2}\;
\;\;\;\;\;\;\;\;\;\;\;\;\;\;\frac{b}{2}\;}
^{\frac{\;\tilde\lambda}{2}-\frac{1}{2b}\;\;\;\;
\frac{\tilde\lambda}{2}+\frac{1}{2b}\;}\right]\!(\,q)\,,
\end{eqnarray}
where the $q$ variable is the elliptic nome introduced in (\ref{nome}).

Notice, that thanks to the AGT relation (\ref{AGTtorus}) and
eq. (\ref{ellipticTorus}), the elliptic torus
block ${\cal H}_{c,\Delta_{\lambda}}^{\tilde\lambda}(q)$
can be replaced with the ${\cal N}=2^{*}$, ${\sf U(2)}$
instanton partition function:
\begin{equation}\label{Hagt}
{\cal H}_{c,\Delta_{\lambda}}^{\tilde\lambda}(q)=
\hat\eta(q)^{2-2\Delta_{\tilde\lambda}}
\,\mathcal{Z}^{{\cal N}=2^{\ast},{\sf U(2)}}_{{\sf
inst}}(q,a,m;\epsilon_1, \epsilon_2),
\end{equation}
where
$$
\lambda=\frac{2a}{\sqrt{\epsilon_1 \epsilon_2}},
\;\;\;\;\;\;\;\;\;\;\;\;
\tilde\lambda=2\sqrt{\frac{m^2}{\epsilon_1 \epsilon_2}+\frac{m}{\sqrt{\epsilon_1 \epsilon_2}}
\left(\frac{\epsilon_1+\epsilon_2}{\sqrt{\epsilon_1 \epsilon_2}}\right)
+\frac{(\epsilon_1 +\epsilon_2)^2}{4\epsilon_1 \epsilon_2}}.
$$

Applying eqs. (\ref{relation}) and (\ref{NSlimit}), one finds
that the identities (\ref{Pogho1}) and (\ref{Hagt}) give
in the limit $b\to 0$ ($\Longleftrightarrow\;\epsilon_2\to 0$):
\begin{eqnarray}
\label{B1}
\frac{1}{\epsilon_1}\;{\cal W}^{{\cal N}=
2^{\ast},{\sf U(2)}}_{{\sf inst}}(q,a,m;\epsilon_1)
- 2\tilde\delta\log\hat\eta(q)
&=&
{\sf h}_{\frac{1}{4}-\frac{a^2}{\epsilon^{2}_{1}}}
\!\left[_{\;\;\;\;\frac{3}{16}\;\;\;\;\;\;
\;\;\;\;\;\;\;\;\;\;\;\;\;\;\;\;\;
\;\;\;\;\;\;\frac{3}{16}}
^{\frac{1}{4}(\frac{3}{4}-\frac{m}{\epsilon_1}(1+\frac{m}{\epsilon_1}))
\;\;\;\;\;
\frac{1}{4}(\frac{3}{4}-\frac{m}{\epsilon_1}(1+\frac{m}{\epsilon_1}))}\right]\!(\,q)\nonumber
\\
\label{B2}
&=&
{\sf h}_{\frac{1}{4}-\frac{a^2}{\epsilon^{2}_{1}}}
\!\left[_{\;\;\;\;\;\frac{1}{4}\;\;\;\;\;\;\;\;\;\;
\;\;\;\;\;\;\;\;\;\;\;\frac{1}{4}}
^{\frac{1}{4}(1-\frac{m^2}{\epsilon_{1}^{2}})\;\;\;\;\;
-\frac{1}{4}\frac{m}{\epsilon_{1}}(2+\frac{m}{\epsilon_{1}})}\right]\!(\,q),
\end{eqnarray}
where
$
{\cal W}^{{\cal N}=
2^{\ast},{\sf U(2)}}_{{\sf inst}}(q,a,m;\epsilon_1)
=\lim_{\epsilon_2 \to 0}\epsilon_2 \log
\mathcal{Z}^{{\cal N}=2^{\ast},{\sf U(2)}}_{{\sf
inst}}(q,a,m;\epsilon_1, \epsilon_2)
$
and the classical conformal weights $\tilde\delta$, $\delta$ and $\delta_i$
(written explicitly in (\ref{B1})) are defined by
\begin{eqnarray*}
\tilde\delta &\equiv&\lim_{b\to 0}b^2 \Delta_{\tilde\lambda}
=\lim_{\epsilon_2\to 0}\frac{\epsilon_2}{\epsilon_1}\Delta_{\tilde\lambda}
=-\frac{m(m+\epsilon_1)}{\epsilon_{1}^{2}},
\\
\delta &\equiv&\lim_{b\to 0}b^2 \Delta_{\lambda}
=\lim_{\epsilon_2\to 0}\frac{\epsilon_2}{\epsilon_1}\Delta_{\lambda}
=\frac{1}{4}-\frac{a^2}{\epsilon_{1}^{2}},
\\
\delta_i &\equiv & \lim_{b\to 0}b^2 \Delta_{\lambda_i}
=\lim_{\epsilon_2\to 0}\frac{\epsilon_2}{\epsilon_1}\Delta_{\lambda_i}.
\end{eqnarray*}

On the other hand, Nekrasov and Shatashvili have found in \cite{NekraSha} that the
instanton twisted superpotential
${\cal W}^{{\cal N}=2^{\ast},{\sf U(N)}}_{{\sf inst}}$
computed from the free energy
$\log\mathcal{Z}_{{\sf inst}}$ of the one-dimensional interacting gas of the
${\cal N}=2^{\ast}$ ${\sf U(N)}$ instanton particles gives the thermodynamic Bethe ansatz \cite{YY}
for the ${\sf N}$-particle eCM system\footnote{$\textrm{Li}_{2}(z)$ is a dilogarithm function defined by
$$
\textrm{Li}_{2}(z)=\sum_{k=1}^{\infty}\frac{z^k}{k^2}
=
\int\limits_{z}^{0}\frac{\log
(1-t)dt}{t}.
$$}:
\begin{equation}\label{rep}
{\cal W}^{{\cal N}=2^{\ast},{\sf U(N)}}_{{\sf inst}}(q,{\sf a},m;\epsilon_1)=\oint\limits_{C}dz
\left[-\frac{1}{2}\,\varphi(z)\log\left(1-q {\cal Q}(z)\textrm{e}^{-\varphi(z)}\right)
+\textrm{Li}_{2}\left(q {\cal Q}(z)\textrm{e}^{-\varphi(z)}\right)\right].
\end{equation}
Here  $\varphi(x)$ is the solution of the integral equation:
\begin{equation}
\label{varphi} \varphi(x)=\oint\limits_{C}dy\,
{\cal G}(x-y)\log\left(1-q {\cal Q}(y)\textrm{e}^{-\varphi(y)}\right).
\end{equation}
The above solution
is determined by the functions ${\cal Q}$, ${\cal G}$ and the contour $C$.
In \cite{NekraSha} it has been found that
\begin{eqnarray}
\label{QG}
{\cal Q}(x)&=&\frac{P(x-m)P(x+m+\epsilon_1)}{P(x)P(x+\epsilon_1)},
\;\;\;\;\;\;\;\;\;\;\;\;\;\;
P(x)=\prod\limits_{i=1}^{{\sf N}}(x-a_i),
\\
{\cal G}(x)&=&\frac{d}{dx}\log\frac{(x+m+\epsilon_1)(x-m)(x-\epsilon_1)}
{(x-m-\epsilon_1)(x+m)(x+\epsilon_1)},
\end{eqnarray}
where $m$ is mass of the
adjoint hypermultiplet and $a_i$, $i=1,\ldots,{\sf N}$ are the
vev's. The contour $C$ on the complex plane comes from
infinity, goes around the points $a_{i}+k\epsilon_1$, $i=1,\ldots,{\sf
N}$, $k=0,1,2,\ldots$ and goes back to infinity. It separates these
points and the points $a_i +lm+k\epsilon_1$, $l\in\mathbb{Z}$,
$k=-1,-2,\ldots\;\;$.

Hence, combining (\ref{Hclassblock}), (\ref{B1}) and (\ref{rep}) for ${\sf N}=2$ one obtains
the following expressions for the two families of the 4-point classical blocks on the sphere:
\begin{eqnarray}\label{BS1}
{\sf f}_{{\frac{1}{4}}-\frac{a^2}{\epsilon^{2}_{1}}}
\!\left[_{\;\;\;\;\frac{3}{16}\;\;\;\;\;\;
\;\;\;\;\;\;\;\;\;\;\;\;\;\;\;\;\;
\;\;\;\;\;\;\frac{3}{16}}
^{\frac{1}{4}(\frac{3}{4}-\frac{m}{\epsilon_1}(1+\frac{m}{\epsilon_1}))
\;\;\;\;\;
\frac{1}{4}(\frac{3}{4}-\frac{m}{\epsilon_1}(1+\frac{m}{\epsilon_1}))}\right]\!(\,x)
&=&
\left(-\frac{1}{8}+\frac{1}{4}\frac{m}{\epsilon_{1}}\left(1+\frac{m}{\epsilon_{1}}\right)\right)\log x
\\
&+&
\left(-\frac{1}{8}+\frac{1}{2}\frac{m}{\epsilon_{1}}\left(1+\frac{m}{\epsilon_{1}}\right)\right)\log (1-x)\nonumber
\\
&+&
2\frac{m}{\epsilon_{1}}\left(1+\frac{m}{\epsilon_{1}}\right)\log(\theta_3(q(x))\,\hat\eta(q(x)))
\nonumber\\
&-&
\frac{a^2}{\epsilon_{1}^{2}}\log(16q(x)) +
\frac{1}{\epsilon_1}\, {\cal W}_{{\sf inst}}^{{\cal N}=2^{\ast},{\sf
U(2)}}(q(x),{\sf a},m;\epsilon_1)\nonumber
\end{eqnarray}
and
\begin{eqnarray}
\label{BS2}
{\sf f}_{\frac{1}{4}-\frac{a^2}{\epsilon^{2}_{1}}}
\!\left[_{\;\;\;\;\;\frac{1}{4}\;\;\;\;\;\;\;\;\;\;
\;\;\;\;\;\;\;\;\;\;\;\frac{1}{4}}
^{\frac{1}{4}(1-\frac{m^2}{\epsilon_{1}^{2}})\;\;\;\;\;
-\frac{1}{4}\frac{m}{\epsilon_{1}}(2+\frac{m}{\epsilon_{1}})}\right]\!(\,x)
&=&
\;
\frac{1}{4}\frac{m}{\epsilon_1}\left(2+\frac{m}{\epsilon_1}\right)\log x
+\frac{1}{2}\frac{m}{\epsilon_1}\left(1+\frac{m}{\epsilon_1}\right)\log(1-x)
\nonumber\\
&+&
2\frac{m}{\epsilon_1}\left(1+\frac{m}{\epsilon_1}\right)\log(\theta_3(q(x))\,\hat\eta(q(x)))
\\
&-&
\frac{a^2}{\epsilon_{1}^{2}}\log(16q(x)) +
\frac{1}{\epsilon_1}\, {\cal W}_{{\sf inst}}^{{\cal N}=2^{\ast},{\sf
U(2)}}(q(x),{\sf a},m;\epsilon_1),\nonumber
\end{eqnarray}
where ${\sf a}=(a_1,a_2)=(a,-a)$. Let us emphasize, that the elliptic
classical blocks, that so far were known only in the form of power series,
in  eqs. (\ref{BS1}) and (\ref{BS2}) are expressed in terms of the eCM Yang's functional.

Notice, that for $m=0$ or $m=-\epsilon_1$, the expressions (\ref{BS1}) and
(\ref{BS2}) give
\begin{eqnarray}\label{ellipClassBlock1}
{\sf f}_{\frac{1}{4}-\frac{a^2}{\epsilon^{2}_{1}}}\!
\left[_{\frac{3}{16}\;\frac{3}{16}}^{\frac{3}{16}\;\frac{3}{16}}\right]\!(\,x)
&=& -\frac{1}{8}\log(x(1-x))-\frac{a^2}{\epsilon^{2}_{1}}\log(16q(x)),\\
\label{ellipClassBlock2}
{\sf f}_{\frac{1}{4}-\frac{a^2}{\epsilon^{2}_{1}}}
\!\left[_{\frac{1}{4}\;\;\;\frac{1}{4}}
^{\frac{1}{4}\;\;\;
0}\right]\!(\,x)
&=&-\frac{a^2}{\epsilon^{2}_{1}}\log(16q(x)),\\
\label{ellipClassBlock3}
{\sf f}_{\frac{1}{4}-\frac{a^2}{\epsilon^{2}_{1}}}
\!\left[_{\frac{1}{4}\;\;\;\frac{1}{4}}
^{0\;\;\;\;
\frac{1}{4}}\right]\!(\,x)
&=&-\frac{1}{4}\log x-\frac{a^2}{\epsilon^{2}_{1}}\log(16q(x)).
\end{eqnarray}
Indeed, if $m\in\lbrace 0, -\epsilon_1\rbrace$
then ${\cal Q}=1$ and ${\cal G}=0$ (cf.(\ref{QG})). As a consequence, $\varphi=0$ and
from (\ref{rep}) we have
$
{\cal W}^{{\cal N}=
2^{\ast},{\sf U(2)}}_{{\sf inst}}=0.
$
This completes our proof.

In order to check the consistency of eqs. (\ref{ellipClassBlock1})-(\ref{ellipClassBlock3})
one can expand the right hand sides making use of
the power expansion of the elliptic nome\footnote{
This expansion can be found from the inverse
$x(q)=\theta_{2}^{4}(q)/\theta_{3}^{4}(q)$
of the nome function where the theta functions are of the form:
\begin{eqnarray*}
\label{theta}
\theta_2 (q) &=& 2q^{\frac{1}{4}}\sum\limits_{n=0}^{\infty} q^{n(n+1)}
= 2q^{\frac{1}{4}}\prod\limits_{n=1}^{\infty}(1-q^{2n})(1+q^{2n})^2,
\\
\theta_3 (q) &=& 1+2\sum\limits_{n=1}^{\infty} q^{n^2}=\prod\limits_{n=1}^{\infty}
(1-q^{2n})(1+q^{2n-1})^2.
\end{eqnarray*}}
\begin{equation}
\label{expNome}
q(x) = {x\over 16} \left( 1 + {x\over 2} + {21 x^2\over 64} +{31 x^3\over 128}+
{6257 x^4\over 32768}+\dots\right).
\end{equation}
Using (\ref{expNome}) one obtains the expansion
\begin{eqnarray*}
-\frac{1}{8}\log(x(1-x))-\frac{a^2}{\epsilon^{2}_{1}}\log(16q(x))&=&
\left(-\frac{1}{8}-\frac{a^2}{\epsilon^{2}_{1}}\right)\log x +\\
&&\hspace{-50pt}+
\left(\frac{1}{8}-\frac{1}{2}\frac{a^2}{\epsilon^{2}_{1}}\right)x
+
\left(\frac{1}{16}-\frac{13}{64}\frac{a^2}{\epsilon^{2}_{1}}\right)x^2
+
\left(\frac{1}{24}-\frac{23}{192}\frac{a^2}{\epsilon^{2}_{1}}\right)x^3\\
&&\hspace{-50pt}+
\left(\frac{1}{32}-\frac{2701}{32768}\frac{a^2}{\epsilon^{2}_{1}}\right)x^4
+
\left(\frac{1}{40}-\frac{31237}{327680}\frac{a^2}{\epsilon^{2}_{1}}\right)x^5 +\ldots,
\end{eqnarray*}
which coincides with the classical block expansion (\ref{classblock})
with $\delta=\frac{1}{4}-\frac{a^2}{\epsilon^{2}_{1}}$
and $\delta_{i}=\frac{3}{16}$. Analogous checks can be made in the case of eqs.
(\ref{ellipClassBlock2}) and (\ref{ellipClassBlock3}).

\section{Uniformization of 4-punctured sphere and the eCM Yang's functional}
The classical Liouville theory is the theory of
the conformal factor $\phi(z,\bar z)$ of the hyperbolic metric on ${\cal
C}_{g,n}$. The conformal factor is a solution to the  Liouville equation:
\begin{equation}
\label{Liouville}
\partial_z\partial_{\bar z} \phi(z,\bar z) =
\frac{\mu}{2}\, \textrm{e}^{\phi(z,\bar z)}.
\end{equation}
The metric on ${\cal C}_{g,n}$ is
determined by the singular behavior of $\phi$ at the punctures $z_1,\ldots,z_n$. Consider
the case of ${\cal C}_{\,g,n}$ being a punctured sphere ${\cal
C}_{\,0,n}$ and choose complex coordinates on ${\cal C}_{\,0,n}$
in such a way that $z_n = \infty.$ The  existence and the uniqueness
of the solution of the equation  (\ref{Liouville}) on the sphere
with the {\it elliptic singularities}, namely
\begin{eqnarray}
\label{asymptot:elliptic}
&&\phi(z,\bar z) = \left\{
\begin{array}{lll}
-2\left(1-\xi_j \right)\log | z- z_j |  + O(1) & {\rm as } & z\to
z_j,
\hskip 5mm j = 1,\ldots,n-1,\\
-2\left(1+\xi_n \right)\log | z| + O(1) & {\rm as } & z\to \infty,
\end{array}
\right.
\\[3pt]
\label{Picard}
&&
\textrm{where}\;\;\xi_i \in \mathbb{R}^{+}\setminus\left\lbrace 0\right\rbrace\;\;\;\textrm{for}\;\textrm{all}\;\;i=1,\ldots,n
\;\;\;\textrm{and}\;\;\sum\limits_{i=1}^{n}\xi_i < n-2,
\end{eqnarray}
were proved by Picard \cite{Picard1,Picard2} (see also
\cite{Troyanov} for a modern proof). That solution can be interpreted
as the conformal factor of the complete, hyperbolic metric on \( {\cal
C}_{\,0,n} = \mathbb{C}\setminus\{z_1,\ldots,z_{n-1}\} \) with
conical singularities characterized by opening angles $\theta_j=2\pi\xi_j$ at
punctures $z_j$. A solution of the Liouville equation is known to exist also in the case
of {\it parabolic singularities} which correspond to $\xi_j \to 0$. In that case
the asymptotic behavior of the Liouville field is:
\begin{equation}
\label{asymptot:parabolic} \phi(z,\bar z) = \left\{
\begin{array}{lll}
-2\log |z- z_j |  -2\log \left|\log |z- z_j |\right| + O(1) & {\rm as } & z\to z_j, \\
-2\log |z| - 2\log \left|\log |z|\right| + O(1) & {\rm as } & z\to
\infty.
\end{array}
\right.
\end{equation}

The central statement in classical Liouville theory is
the famous {\it uniformization theorem} firstly proved by
Poincar\'{e} and Koebe (1907). The uniformization theorem states that every Riemann surface ${\cal C}$
is conformally  equivalent:
\begin{quote}
(i) to the Riemann sphere $\mathbb{C}\cup\lbrace\infty\rbrace$;

\noindent
(ii) or to the upper half plane $\mathbb{H}=\lbrace \tau\in\mathbb{C}\,:\,\mathfrak{Im}\,\tau
> 0\rbrace$;

\noindent
(iii) or to a quotient of $\mathbb{H}$ by a discrete subgroup
$G\subset {\sf PSL}(2,\mathbb{R})\equiv {\sf SL}(2,\mathbb{R})/\mathbb{Z}_2$
acting as M\"{o}bius transformations.
\end{quote}
From this theorem it follows the existence
of a meromorphic function
$$
\lambda\! : \mathbb{H}\ni\tau\to
z=\lambda(\tau)\in\mathbb{H}\diagup G\cong{\cal C},
$$
called {\it uniformization}. The map $\lambda$ is explicitly known
only for the 3-punctured sphere \cite{Ahlfors} and in a few very special,
symmetric cases with higher number of punctures \cite{Hempel}. In
particular, an explicit construction of this map for the 4-punctured
sphere is a longstanding and still open problem.

One possible method to construct $\lambda$ in the case of the $n$-punctured
sphere ${\cal C}_{\,0,n}$ with parabolic singularities
has been proposed by Poincar\'{e}. This construction is based on the relation of the
uniformization problem to a certain Fuchs equation on ${\cal C}_{\,0,n}$.
If $\lambda$ is the uniformization of ${\cal C}_{\,0,n}$, the inverse map
$$
\omega=\lambda^{-1}\! :
{\cal C}_{\,0,n}\ni z\to \tau(z)\in\mathbb{H}
$$
is a multi-valued function with branch points $z_j$ and with branches related by
the elements $T_k$ of the group $G$. One can show that the Schwarzian
derivative of $\omega$ is a holomorphic function on ${\cal
C}_{\,0,n}$ of the form \cite{Hempel}:
\begin{eqnarray}
\label{Schuni}
\left\lbrace\omega,  z\right\rbrace &=&
\sum_{k=1}^{n-1}\left(\frac{\frac{1}{2}}{(z - z_k)^2}+
\frac{2 c_k}{z - z_k}\right),\\
\left\lbrace\omega,  z\right\rbrace &\stackrel{z\to\infty}{=}&
\frac{\frac{1}{2}}{z^2}+{\cal O}(z^{-3}),
\end{eqnarray}
where the so-called {\it accessory parameters} $c_k$ satisfy the
relations
\begin{equation}
\label{dodatki}
\sum_{k=1}^{n-1}c_k =0, \;\;\;\;\;\;\;\;\;\;
\;\;\;\;\;\;\;\;\;\;
\sum_{k=1}^{n-1}\left(4c_k z_k + 1 \right)=1,
\end{equation}
On the other hand it is a well known fact \cite{Hempel,TZ}
that if $\lbrace\psi_1, \psi_2\rbrace$ is a fundamental system of
normalized ($\psi_1 \psi_{2}' - \psi_{1}'\psi_{2}=1$) solutions of the
Fuchs equation
\begin{equation}\label{Fuchs}
\psi(z)'' + {\textstyle\frac{1}{2}}\,\lbrace \omega,
z\rbrace\,\psi(z)=0
\end{equation}
with ${\sf SL}(2, \mathbb{R})$ monodromy with respect to all
punctures then up to a M\"{o}bius transformation the inverse map is
$$
\omega=\frac{\psi_{1}}{\psi_{2}}\;.
$$
Therefore, it is possible to reformulate the uniformization problem of ${\cal C}_{\,0,n}$
as a kind of Riemann-Hilbert problem for the Fuchs equation
(\ref{Fuchs}), where the elements $T_k \in G$ correspond
to monodromy matrices $M_k$ around each puncture $z_k$. An applicability
of the Fuchs equation to the calculation of the inverse map $\omega$ depends
on our ability to calculate the accessory parameters and to choose
an appropriate fundamental system of normalized solutions.
In the case of three punctures the accessory parameters
are determined by the equations (\ref{dodatki}).
In the case in which $n>3$ the Liouville theory on ${\cal C}_{\,0,n}$ becomes helpful,
because eqs. (\ref{dodatki}) do not provide enough constraints in order to calculate the $c_k$'s.

The connection with LFT comes out from the existence
of the Poincar\'{e}-Klein metric $
d\textrm{s}^{2}_{\mathbb{H}}=d\tau\,d\bar\tau/\left(
\mathfrak{Im}\,\tau\right)^2$ on the upper half plane $\mathbb{H}$. The pull back:
$$
\omega^{\ast}d\textrm{s}^{2}_{\mathbb{H}}=
\frac{1}{\left( \mathfrak{Im}\,\tau\right)^2}\left|
\frac{\partial\tau}{\partial z}\right|^2 \ dz d\bar
z = \textrm{e}^{\phi(z,\bar z)} dz d\bar z
$$
is a regular hyperbolic metric on ${\cal C}_{\,0,n}$, conformal to
the standard flat matric $dz d\bar z$ on
$\mathbb{C}$.
The conformal factor $\phi(z, \bar z)$
satisfies the Liouville equation on ${\cal C}_{\,0,n}$ with the
asymptotic condition (\ref{asymptot:parabolic}). One can show
that the energy-momentum tensor $T(z)$ of this solution is equal to
one half of the Schwarzian derivative of the inverse map:
\begin{equation}\label{Schwarz}
T(z) \equiv - {\textstyle\frac{1}{4}}\,(\partial_{z}\phi)^2 +
{\textstyle\frac{1}{2}}\,\partial_{z}^{2}\phi
={\textstyle\frac{1}{2}}\,\lbrace\omega,  z\rbrace.
\end{equation}
This allows to calculate all accessory parameters once the classical
solution $\phi$ is known.
However, the problem of finding solutions of
the Liouville equation seems to be at least as hard as the problem
of constructing the map $\omega$.
For this reason, the application of Liouville
theory at this level does not lead to an essential simplification and is not
much helpful.

So far we discussed the uniformization problem of the $n$-punctured sphere
with parabolic singularities. However, one can consider a more general
problem of the sphere with $n$ elliptic singularities characterized by real parameters $\xi_j > 0$.
The notion of accessory parameters can be introduced in terms of the energy-momentum
tensor of the solution $\phi(z,\bar z)$ with the asymptotic condition
(\ref{asymptot:elliptic}). In the present case the energy-momentum
tensor is of the form \cite{TZ}:
\begin{eqnarray}
\label{TensorElliptic1}
T(z) &=&
\sum_{k=1}^{n-1}\left(\frac{\delta_k}{(z - z_k)^2} +
\frac{c_k}{z - z_k}\right),
\\
\label{TensorElliptic2}
T(z) &\stackrel{z\to\infty}{=}&
\frac{\delta_n}{z^2}+{\cal O}(z^{-3}).
\end{eqnarray}
The accessory parameters $c_k$ obey
\begin{equation}
\label{dodatki2}
\sum_{k=1}^{n-1}c_k =0, \;\;\;\;\;\;\;\;\;\;
\;\;\;\;\;\;\;\;\;\;
\sum_{k=1}^{n-1}\left(\delta_k + c_k z_k\right)=\delta_n,
\end{equation}
and as before they are not fully determined for $n>3$.
In eqs. (\ref{TensorElliptic1}), (\ref{TensorElliptic2}) and (\ref{dodatki2})
the parameters $\delta_j=\frac{1}{4}(1-\xi_{j}^{2})$, $j=1,\ldots,n$ are
the classical conformal weights. The multi-valued function
$\omega$ is still of interest.
Having $T(z)$ given by (\ref{TensorElliptic1}) one can consider the Fuchs equation
\begin{equation}\label{Fuchs2}
\psi(z)'' + T(z)\,\psi(z) = 0,
\end{equation}
which monodromy group is a subgroup of ${\sf PSL}(2,\mathbb{R})$.
The classical result (\ref{Schwarz}) holds,
where $\omega=\psi_1/\psi_2$ is the ratio of two linearly independent solutions $\psi_1$, $\psi_2$
of the eq. (\ref{Fuchs2}).
Then, the map $\omega$ can be computed if one can calculate the accessory parameters in the eq.
(\ref{Fuchs2}) and select fundamental solutions with a suitable monodromy.

For almost a century the problem of accessory parameters has been
unsolved until the appearance of the solution proposed by Polyakov
(as reported in refs. \cite{TZ,Tak,TZ2}).
The so-called Polyakov conjecture states that the properly defined
and normalized Liouville action functional evaluated on the
classical solution $\phi(z, \bar z)$ is the generating functional for
the accessory parameters:
\begin{equation}
\label{PC} c_j = -\frac{\partial S_{\sf L}^{\,\sf
cl}[\phi]}{\partial z_j}.
\end{equation}
This formula was derived within the path integral approach to the
quantum Liouville theory by analyzing the quasi-classical limit of the
conformal Ward identity \cite{Tak}. In the case of parabolic
singularities on the $n$-punctured Riemann sphere a rigorous proof has been
given by Zograf and Takhtajan \cite{TZ2}. Alternative proofs, valid
both in the case of parabolic and general elliptic singularities,
have been proposed in \cite{CMS,TZ}. The Polyakov conjecture can be
proved also in the case of the so-called hyperbolic singularities
representing ``holes'' with geodesic boundaries \cite{HJ1}.

The next essential step paving the way for an explicit uniformization
of the 4-punctured sphere was done by brothers Zamolodchikov
\cite{Zamolodchikov:1995aa}. Studying the classical limit of the
4-point function of the quantum Liouville theory they argued that
the classical Liouville action on a sphere with four elliptic (parabolic)
singularities can be expressed in terms of the classical Liouville
action for three singularities and the 4-point classical block.

Indeed, the 4-point function of the Liouville primary operators
located at $\infty, 1, x, 0$ is expressed as an integral over the
continuous spectrum of the theory
of $s$-channel conformal blocks and DOZZ 3-point functions:
\begin{eqnarray}
\label{four:point:} && \hspace*{-2.5cm} \Big\langle
\textsf{V}_{\alpha_4}(\infty,\infty)\textsf{V}_{\alpha_3}(1,1)
\textsf{V}_{\alpha_2}(x,\bar x)\textsf{V}_{\alpha_1}(0, 0)
\Big\rangle =
\\
\nonumber && \int\limits_{\frac{Q}{2} + i
\mathbb{R}^{+}}\!\!\!\!\!\!\! d\alpha\;
C(\alpha_4,\alpha_3,\alpha)C(Q-\alpha,\alpha_2,\alpha_1) \left|
{\cal F}_{\!1+6Q^2,\Delta}
\!\left[_{\Delta_{4}\;\Delta_{1}}^{\Delta_{3}\;\Delta_{2}}\right]
\!(x) \right|^2.
\end{eqnarray}
Let
$$
\mbox{\bf 1}_{\Delta,\Delta} = \sum_{I}
(|{\nu_{\Delta,I}}\rangle\otimes |{\nu_{\Delta,I}}\rangle)
(\langle{\nu_{\Delta,I}}|\otimes\langle{\nu_{ \Delta,I}}|)
$$
be the operator that projects onto the space spanned by the states
belonging to the conformal family with the highest weight $\Delta$. The
correlation function with the $\mbox{\bf 1}_{\Delta,\Delta}$
insertion factorizes into a product of holomorphic and
anti-holomorphic factors:
\begin{eqnarray}
\label{c4} && \hspace*{-1cm}
 \Big\langle
\textsf{V}_{\alpha_4}(\infty,\infty)\textsf{V}_{\alpha_3}(1,1)
\mbox{\bf 1}_{\Delta,\Delta}
\textsf{V}_{\alpha_2}(x,\bar x)\textsf{V}_{\alpha_1}(0, 0)
\Big\rangle  =  \\
 \nonumber
&& C(\alpha_4,\alpha_3,\alpha)\,C(Q-\alpha,\alpha_2,\alpha_1)\,
{\cal F}_{\!1+6Q^2,\Delta}
\!\left[_{\Delta_{4}\;\Delta_{1}}^{\Delta_{3}\;\Delta_{2}}\right]
\!(x)\, {\cal F}_{\!1+6Q^2,\Delta}
\!\left[_{\Delta_{4}\;\Delta_{1}}^{\Delta_{3}\;\Delta_{2}}\right]
\!(\bar x).
\end{eqnarray}
Assuming a path integral representation for the l.h.s.
and heavy conformal weights $\Delta,\Delta_i \sim \frac{1}{b^2}$
one should expect in the limit $b \to 0$ the  following asymptotic behavior
\begin{equation}
\label{a4} \Big\langle
\textsf{V}_{\alpha_4}(\infty,\infty)\textsf{V}_{\alpha_3}(1,1)
\mbox{\bf 1}_{\Delta,\Delta}
\textsf{V}_{\alpha_2}(x,\bar x)\textsf{V}_{\alpha_1}(0, 0)
 \Big\rangle \sim
{\rm e}^{-\frac{1}{b^2} S_{\sf L}^{\sf cl}(\delta_i,x;\delta) }.
\end{equation}
$S_{\sf L}^{\,\sf cl}(\delta_i,x;\delta)$ is the 4-point classical
Liouville action, i.e. the properly defined Liouville action functional
evaluated on the classical solution of the Liouville equation
(\ref{Liouville}) on ${\cal C}_{\,0,4}$ with the asymptotic conditions
(\ref{asymptot:parabolic}) or (\ref{asymptot:elliptic}). The
parameters $\delta$, $\delta_i$, $i=1,\ldots,4$ are the
classical conformal weights defined by
$\Delta=\frac{1}{b^2}\,\delta$ and $\Delta_i=\frac{1}{b^2}\,\delta_i$, where
$\delta,\delta_i = {\cal O}(1)$. On the other hand, one can calculate
this limit for the DOZZ coupling constants obtaining
\cite{Zamolodchikov:1995aa,Hadasz:2003he}
\begin{equation}
\label{asymptotC}
C(\alpha_4,\alpha_3,\alpha)C(Q-\alpha,\alpha_2,\alpha_1)
 \sim
{\rm e}^{-\frac{1}{b^2}\left( S_{\sf L}^{\sf
cl}(\delta_4,\delta_3,\delta) + S_{\sf L}^{\sf
cl}(\delta,\delta_2,\delta_1) \right)}.
\end{equation}
$S_{\sf L}^{\,\sf cl}(\delta_1,\delta_2,\delta)$ is the 3-point
classical Liouville action which in the case of two parabolic
or elliptic weights given respectively by $\delta_1=\delta_2=\frac{1}{4}$
and $\delta_i=\frac{1}{4}\,(1-\xi_{i}^2)$, $i=1,2$
and one hyperbolic weight $\delta=\frac{1}{4}+p^2$ with $p\in\mathbb{R}$
reads as follows \cite{Zamolodchikov:1995aa,Hadasz:2003he}:
\begin{eqnarray}
\label{action:mixed:3} \nonumber S_{\sf L}^{\,\sf
cl}(\delta_1,\delta_2,\delta) & = & \frac12(1-\xi_1-\xi_2)\log\mu +
\!\!\!\!\!\sum\limits_{\sigma_2,\sigma_3 = \pm}\!\!\!\!
F\left(\frac{1-\xi_1}{2} + \sigma_2\frac{\xi_2}{2} +i
\sigma_3p\right)
\\
&&-\sum\limits_{j=1}^2F(\xi_j)
 + H(2i p) + \pi|p| +{\rm const},
\end{eqnarray}
where
\[
F(x)  =  \int\limits_{1/2}^{x}\!dy\;
\log\frac{\Gamma(y)}{\Gamma(1-y)}, \;\;\;\;\;\;\;\;\;\;\;\;\;\;\;\
H(x)  =
\int\limits_0^x\!dy\;\log\frac{\Gamma(-y)}{\Gamma(y)}.
\]
It follows that for $b\to 0$ the conformal block should have the
exponential behavior (\ref{defccb}). Hence
\begin{equation}
\label{deltaaction} S_{\sf L}^{\,\sf cl}(\delta_i,x;\delta)=S_{\sf
L}^{\;\sf cl}(\delta_4,\delta_3,\delta) + S_{\sf L}^{\;\sf
cl}(\delta,\delta_2,\delta_1)- {\sf f}_{\delta}
\!\left[_{\delta_{4}\;\delta_{1}}^{\delta_{3}\;\delta_{2}}\right](x)
-\bar {\sf f}_{\delta}
\!\left[_{\delta_{4}\;\delta_{1}}^{\delta_{3}\;\delta_{2}}\right](\bar
x).
\end{equation}

In the semiclassical limit the l.h.s.
of formula (\ref{four:point:}) takes the form ${\rm
e}^{-\frac{1}{b^2} S_{\sf L}^{\,\sf
cl}(\delta_4,\delta_3,\delta_2,\delta_1;x)}$, where
$$
S_{\sf L}^{\,\sf cl}(\delta_4,\delta_3,\delta_2,\delta_1;x) \equiv
S_{\sf L}^{\,\sf
cl}(\delta_4,\delta_3,\delta_2,\delta_1;\infty,1,x,0) .
$$
The r.h.s. of (\ref{four:point:}) is in this limit determined by the
saddle point approximation
$$
 {\rm e}^{-\frac{1}{b^2} S_{\sf L}^{\,\sf cl}(\delta_i,x)}
\; \approx \;
 \int\limits_0^\infty\!dp\; {\rm e}^{-\frac{1}{b^2} S_{\sf L}^{\,\sf cl}(\delta_i,x;\delta)}
$$
where $ \delta ={\textstyle {1\over 4}} +p_{s}(x)^2 $ and the
saddle point Liouville momentum $p_s(x)$ is determined by
\begin{equation}
\label{saddle} {\partial \over \partial p} S_{\sf L}^{\,\sf cl}
(\delta_i,x;{\textstyle {1\over 4}} +p^2)_{|p=p_s}=0\ .
\end{equation}
One thus gets the factorization
\begin{eqnarray}
\label{clasfact} S_{\sf L}^{\,\sf
cl}(\delta_4,\delta_3,\delta_2,\delta_1;x) &=& S_{\sf L}^{\,\sf
cl}(\delta_4,\delta_3,\delta_s(x)) +
S_{\sf L}^{\,\sf cl}(\delta_s(x),\delta_2,\delta_1)\nonumber\\[10pt]
&-&\,{\sf f}_{\delta_s(x)}
\!\left[_{\delta_{4}\;\delta_{1}}^{\delta_{3}\;\delta_{2}}\right](x)
-\bar {\sf f}_{\delta_s(x)}
\!\left[_{\delta_{4}\;\delta_{1}}^{\delta_{3}\;\delta_{2}}\right](\bar
x)
\end{eqnarray}
first obtained in \cite{Zamolodchikov:1995aa} and
as its consistency condition\footnote{The
classical bootstrap equations have been numerically verified for
punctures \cite{HJP} and for punctures and up to two elliptic
singularities \cite{P}.} the {\it classical
bootstrap equations}:
\begin{eqnarray}
\nonumber
&& \hspace{-80pt} S_{\sf L}^{\,\sf
cl}(\delta_4,\delta_3,\delta_s(x))
+
S_{\sf L}^{\,\sf
cl}(\delta_s(z),\delta_2,\delta_1)
-\,{\sf f}_{\delta_s(x)}
\!\left[_{\delta_{4}\;\delta_{1}}^{\delta_{3}\;\delta_{2}}\right](x)
-\bar {\sf f}_{\delta_s(x)}
\!\left[_{\delta_{4}\;\delta_{1}}^{\delta_{3}\;\delta_{2}}\right](\bar x)
\\[10pt]
\label{clasboot}
&=&
S_{\sf L}^{\,\sf
cl}(\delta_4,\delta_1,\delta_t(x))
+
S_{\sf L}^{\,\sf
cl}(\delta_t(x),\delta_2,\delta_3)\\
\nonumber
&&
\hspace{70pt}-\,{\sf f}_{\delta_t(x)}
\!\left[_{\delta_{4}\;\delta_{3}}^{\delta_{1}\;\delta_{2}}\right](1-x)
-\bar {\sf f}_{\delta_t(x)}
\!\left[_{\delta_{4}\;\delta_{3}}^{\delta_{1}\;\delta_{2}}\right](1-\bar x)
\\[10pt]
\nonumber
&=&  2\delta_2 \log x\bar x + S_{\sf L}^{\,\sf
cl}(\delta_1,\delta_3,\delta_u(x))
+
S_{\sf L}^{\,\sf
cl}(\delta_u(x),\delta_2,\delta_4)\\
\nonumber
&&
\hspace{70pt}-\,{\sf f}_{\delta_u(x)}
\!\left[_{\delta_{1}\;\delta_{4}}^{\delta_{3}\;\delta_{2}}\right]\left({1\over x}\right)
-\bar {\sf f}_{\delta_u(x)}
\!\left[_{\delta_{1}\;\delta_{4}}^{\delta_{3}\;\delta_{2}}\right]\left({1\over \bar x}\right),
\\
&&\delta_t(x) = \delta_s(1-x),\;\;\;\;\;\;\delta_u(x) = \delta_s\left({1\over x}\right).\nonumber
\end{eqnarray}

Turning to the problem of uniformization of the 4-punctured sphere one
sees, that having (\ref{deltaaction}) one can apply the Polyakov
conjecture (\ref{PC}) and calculate the accessory parameter of the
appropriate Fuchsian equation (cf.\cite{HJuniformizacja}).
Surprisingly, the twisted superpotential/Yang's functional appears
in that context.

According to (\ref{Hclassblock}) and (\ref{B1}) we have
\begin{eqnarray}
\label{clW}
{\sf f}_{\delta}\!\left[_{\delta_{4}\;\delta_{1}}^{\delta_{3}\;\delta_{2}}\right]\!(\,x)
&=&
 ({\textstyle {1\over 4}}-\delta_1-\delta_2) \log x
+
 ({\textstyle {1\over 4}}-\delta_2-\delta_3) \log (1-x)\\
 \nonumber
&+&
 (3-4(\delta_1+\delta_2+\delta_3+\delta_4)) \log\theta_3(q(x))\\
 \nonumber
 &+& (\delta -{\textstyle {1\over 4}}) \log (16q(x))
+2\frac{m(m+\epsilon_1)}{\epsilon^{2}_{1}}\log\hat\eta(q(x))\\
&+& \frac{1}{\epsilon_1}\,
{\cal W}_{{\sf inst}}^{{\cal N}=2^{\ast},{\sf
U(2)}}(q(x),a,m;\epsilon_1),\nonumber
\end{eqnarray}
where $\delta=\frac{1}{4}-\frac{a^2}{\epsilon^{2}_{1}}$ and
\begin{equation}
\label{w1}
\delta_{4}=\delta_{1}={\textstyle\frac{3}{16}}={\textstyle\frac{1}{4}}\,(1-({\textstyle\frac{1}{2}})^2),
\;\;\;\;\;\;\;\;\;\;\;\;
\delta_{3}=\delta_{2}={\textstyle\frac{1}{4}}\,(1-\xi^2),
\;\;\;\;\;\;\;\;
\xi=\sqrt{{\textstyle\frac{1}{4}}+\frac{m}{\epsilon_1}(1+\frac{m}{\epsilon_1})},
\end{equation}
or
\begin{equation}
\delta_{4}=\delta_{1}=\frac{1}{4},
\;\;\;\;\;\;\;\;\;\;
\delta_3=\frac{1}{4}\left(1-\frac{m^2}{\epsilon_{1}^{2}}\right),
\;\;\;\;\;\;\;\;
\delta_2=-\frac{1}{4}\,\frac{m}{\epsilon_{1}}\left(2+\frac{m}{\epsilon_{1}}\right).
\end{equation}

Consider the classical block (\ref{clW}) with the weights given by (\ref{w1}).
It can be used to construct the 4-point classical Liouville action:
$$
S_{\sf L}^{\,\sf cl}(\delta_i,x)=S_{\sf L}^{\,\sf cl}
({\textstyle\frac{3}{16}},
{\textstyle\frac{1}{4}}\,(1-\xi^2),
{\textstyle\frac{1}{4}}\,(1-\xi^2),
{\textstyle\frac{3}{16}},x;{\textstyle\frac{1}{4}}+p_{s}(x)^2),
\;\;\;\;\;\;\;\;
0<\xi<{\textstyle\frac{1}{2}},
$$
corresponding to the classical solution on ${\cal C}_{0,4}$
with four elliptic singularities characterized by
$\xi_4=\xi_1=\frac{1}{2}$ and $\xi_3=\xi_2 =\xi$.
The condition $0<\xi<{\textstyle\frac{1}{2}}$, which is fulfilled for
$\frac{m}{\epsilon_1}\in(-1,0)\setminus\lbrace-\frac{1}{2}\rbrace$, ensures
the existence of such solution\footnote{Unfortunately, the
configuration of the four elliptic singularities
with $\xi_4=\xi_3=\xi_2=\xi_1=\frac{1}{2}$
for which we have found the classical block (\ref{ellipClassBlock1})
does not fulfil Picard's  inequality (\ref{Picard}).}
(cf.(\ref{Picard})). The saddle point momentum
$p_{s}(x)$ is determined by the equation (\ref{saddle}), where
$p=i a/\epsilon_{1}$. The Polyakov conjecture in the case under
consideration reads
\begin{eqnarray*}
c_{2}(x)&=&-\frac{\partial}{\partial x}S_{\sf L}^{\,\sf cl}(\delta_i,x)
\\[5pt]
&=&
-\frac{\partial}{\partial p}
S_{\sf L}^{\,\sf cl}(\delta_i,x,{\textstyle\frac{1}{4}}+p^2)\Big|_{p=p_{s}(x)}
\;\frac{\partial p_{s}(x)}{\partial x}-\frac{\partial}{\partial x}
S_{\sf L}^{\,\sf cl}(\delta_i,x,{\textstyle\frac{1}{4}}+p^2)\Big|_{p=p_{s}(x)}
\\[5pt]
&=&
-\frac{\partial}{\partial x}
S_{\sf L}^{\,\sf cl}(\delta_i,x,{\textstyle\frac{1}{4}}+p^2)\Big|_{p=p_{s}(x)}
=
\frac{\partial}{\partial x}\,{\sf f}_{\frac{1}{4}+p^2}
\!\left[_{\;\frac{3}{16}\;\;\;\;\;\;\;\;\;\;\;\;\;
\;\;\;\;\frac{3}{16}\;}
^{\frac{1}{4}(1-\xi^2)
\;
\frac{1}{4}(1-\xi^2)}\right]\!(x)\Big|_{p=p_{s}(x)}
\\[5pt]
&=&
\frac{4\pi K(1-x)p_{s}(x)^2 - 4\pi E(1-x)p_{s}(x)^2-4\xi^2 E(x)+E(x)}
{8(x-1)x K(x)}
\\
&-&\frac{\pi E(x)K(1-x)p_{s}(x)^2}{2(x-1) x K(x)^2}
-\frac{4x-1-4\xi^2}{16(x-1)x}
\\[5pt]
&+&
\frac{2m(m+\epsilon_1)}{\epsilon^{2}_{1}}
\frac{\partial}{\partial x}
\log\hat\eta(q(x))
+
\frac{1}{\epsilon_1}\,
\frac{\partial}{\partial x}\,
{\cal W}_{{\sf inst}}^{{\cal N}=2^{\ast},{\sf
U(2)}}(q(x),a,m;\epsilon_1)\Big|_{\frac{ia}{\epsilon_1}=p_{s}(x)},
\end{eqnarray*}
where (\ref{deltaaction}), (\ref{action:mixed:3}) and (\ref{clW}), (\ref{w1}) have been used.
The symbol $E(x)$ above denotes the complete elliptic integral of the second kind.
This calculation shows an amazing relation between the accessory parameter
and the instanton part of the ${\cal N}=2^{\ast},{\sf U(2)}$ twisted superpotential/Yang's functional.
This calculation reveals also the relation between the vev $a$ and the so-called {\it geodesic
length functions} (cf.\cite{DGOT}):
$$
a=-i\epsilon_1\, p_{s}(x)=-i\frac{\epsilon_1\sqrt{\mu}}{4\pi}\,\ell_{s}(\,x).
$$

Indeed,  let us recall that the classical solution $\phi(z, \bar z)$
on $ {\cal C}_{\,0,n}$ describes a unique hyperbolic geometry
with singularities at the locations of conformal weights. For
 elliptic, parabolic and hyperbolic weights one gets
conical singularities, punctures and holes with geodesic boundaries,
respectively \cite{sei,Hadasz:2003he}.
In the latter case the (classical) conformal weight $\delta$ is related to the length
$\ell$ of the corresponding
hole by
\begin{equation}
\label{glength} \delta =\frac{1}{4}+
\frac{\mu}{4}\left(\frac{\ell}{2\pi}\right)^2,
\end{equation}
where the scale of the classical configuration is set by the condition $R=-{\mu\over 2}$
imposed on the constant scalar curvature $R$.
In the case of 4 singularities at the standard locations
$0,x,1,\infty$ there are  three closed ge\-o\-de\-sics $\Gamma_s,
\Gamma_t, \Gamma_u$ separating the singular points into pairs
$(x,0|1,\infty),$ $(x,1|0,\infty)$ and $(x,\infty|0,1)$
respectively. Since the spectrum of DOZZ theory is hyperbolic, the
singularities corresponding to the saddle point weights
$\delta_i(x)$ are geodesic holes. One may expect that these
weights are related to the lengths $\ell_i$ of the closed
geodesics $\Gamma_i$ in the corresponding channels:
\begin{equation}
\label{glengthII} \delta_i(x) =\frac{1}{4}+
\frac{\mu}{4}\left(\frac{\ell_i(x)}{2\pi}\right)^2,\;\;\;i=s,t,u.
\end{equation}
As a final remark let us point out that the conjectured formula (\ref{glengthII})
has strong numerical support, cf.\cite{HJP}.

\section{Conclusions}
In this work we have derived expressions for the two families of the 4-point classical blocks on the sphere.
We have found, that the most non-trivial parts of the classical blocks are encoded by the Nekrasov-Shatashvili formula for the
instanton part of the ${\cal N}=2^{\ast}$ ${\sf U(2)}$ twisted superpotential/eCM Yang's functional.
Next, we have used one of these new expressions of the classical blocks to calculate
the accessory parameter of the Fuchsian uniformization of the 4-punctured sphere and have found its relation to
${\cal W}_{{\sf inst}}^{{\cal N}=2^{\ast},{\sf U(2)}}$.
Thirdly, we have established a relationship between the 4-point classical block on the sphere
and the $N_f=4$ ${\sf SU(2)}$ (or ${\sf U(2)}$)
twisted superpotential and further used this relationship
to re-derive the Seiberg-Witten prepotential from the classical block.

There are possible extensions of the present work.
Firstly, it would be interesting to apply
the AGT duality and the Bethe/gauge correspondence to learn something new about the matrix models.
Secondly, one can continue the studies of the triple correspondence:
2dCFT/${\cal N}=2$ gauge theories/quantum integrable systems
going beyond ${\sf SU(2)}$ theories, i.e.
looking at the generalization of the AGT duality to
the correspondence between 2d conformal Toda and 4d ${\cal N}=2$ ${\sf
SU(N)}$ gauge theories \cite{Wyllard,MMU(3)}.

As a final remark let us stress that it seems to be an interesting
task to study possible overlaps of our results and those in papers
\cite{Teschner,NekraRosSha}.

\section*{Acknowledgments}
The author is grateful to Paulina Suchanek, Franco Ferrari, Evgeny Ivanov
and Zbigniew Jask\'{o}lski for stimulating discussions and very valuable advices.
Special thanks go to Artur Pietrykowski for assistance at the initial stage of
this work.

The author is also grateful to the organizers of the workshop:
{\it Branes and Bethe Ansatz in Supersymmetric Gauge Theories}
(Simons Center for Geometry And Physics, Stony Brook, March 2011)
for the invitation and for the opportunity to present results
collected in this work.

The author would like to thank the University of
Szczecin and the Faculty of Mathematics and Physics of that
University for the kind hospitality.

\end{document}